# Spectral Oscillations and Temporal Meandering of Dissipative Solitons in a Moving Intracavity Phase Potential

*Heng Dong, Xiankun Yao* and Wen-Li Yang*

H Dong, X Yao
School of Physics, Northwest University, Xi'an 710127, China
E-mail: yaoxk@nwu.edu.cn

X Yao, W Yang
Shaanxi Key Laboratory for Theoretical Physics Frontiers, Xi'an 710127, China

X Yao
Fundamental Discipline Research Center for Quantum Science and Technology of Shaanxi Province, Xi'an 710127, China

W Yang
Institute of Modern Physics, Northwest University, Xi'an 710127, China
Peng Huanwu Center for Fundamental Theory, Xi'an 710127, China

Funding: National Natural Science Foundation of China (NSFC) (Grant No. 12475006, 12247103, and 12434006).

**Abstract:** Dissipative solitons (DSs) provide a prototypical platform for studying nonequilibrium dynamics governed by the interplay between nonlinearity, dissipation, and external driving. In particular, subjected to external potentials, DSs exhibit a rich variety of nonlinear dynamical behaviors. This work investigates DS dynamics in a moving intracavity potential generated by phase modulation whose frequency is detuned from an integer multiple of the cavity repetition rate. Modulation detuning introduces a timescale mismatch between the DS and the moving potential, leading to two distinct dynamical regimes. At large detuning, the DS cannot continuously track the moving potential and exhibits persistent spectral oscillations and temporal meandering, whereas at sufficiently small detuning it enters a synchronized locking state. Furthermore, both the oscillation period and amplitude can be tuned via the modulation detuning and modulation depth. These results reveal how asynchronous driving reshapes DS dynamics and establish a route for controlling their dynamical response in nonlinear optical cavities.

## 1.Introduction

Dissipative solitons (DSs) are typical localized coherent structures in nonequilibrium nonlinear systems, [1–4] arising from the balance among gain, loss, dispersion, and nonlinearity. [5,6] They have found promising applications in waveguide arrays, [7-10] precision material processing, [11] and integrated photonic circuits. [12,13] Among various experimental platforms, mode-locked fiber lasers, owing to their highly controllable intracavity conditions and well-defined dissipative environment, have become ideal systems for investigating DS dynamics. [5,14-20] In such systems, DSs exhibit not only rich stationary states and transient dynamical behaviors, [21,22] but also strong sensitivity to external driving perturbations, making them effective probes for exploring driven nonlinear dynamics in intracavity systems. [23-28] In particular, intracavity phase modulation can directly manipulate the phase, spectral, and temporal dynamics of DSs, [29,30] providing a powerful route for controlling optical-field evolution in driven dissipative media.

Intracavity phase modulation has long been used to control soliton dynamics in optical fiber systems. In conservative fiber systems, early studies showed that weak phase perturbations can reshape soliton timing dynamics and produce drift-balance behavior under external forcing, enabling trapping and synchronization phenomena. [31-33] More recently, phase-modulation techniques have been applied to dissipative solitons in mode-locked fiber lasers, [29,34–37] where soliton dynamics are governed by a stationary phase landscape in the cavity frame. In contrast, modulation detuning introduces a continuously drifting phase potential relative to the intracavity field, [38–42] driving dissipative solitons into intrinsically nonstationary dynamical behavior. The dynamical consequences of such a moving phase potential in dissipative soliton systems remain largely unexplored.

In this work, we investigate the spectral–temporal dynamics of DSs in a moving intracavity phase potential generated by asynchronous phase modulation in a mode-locked fiber laser. By combining experiments, numerical simulations, and theoretical analysis, we show how modulation detuning governs the observed dynamical regimes and gives rise to detuning-dependent oscillatory dynamics, while modulation depth further shapes these dynamics. This detuning introduces a timescale mismatch between the DS and the moving phase potential, providing a clear physical perspective for understanding the observed dynamics. A reduced theoretical model captures the key features of the observed spectral–temporal dynamics and determines the conditions for the emergence of a stationary frequency-shifted state. These results provide physical insight into the response of DSs to asynchronous driving and establish a framework for predicting and controlling their oscillatory dynamics.

## 2. Principle and Theoretical Framework

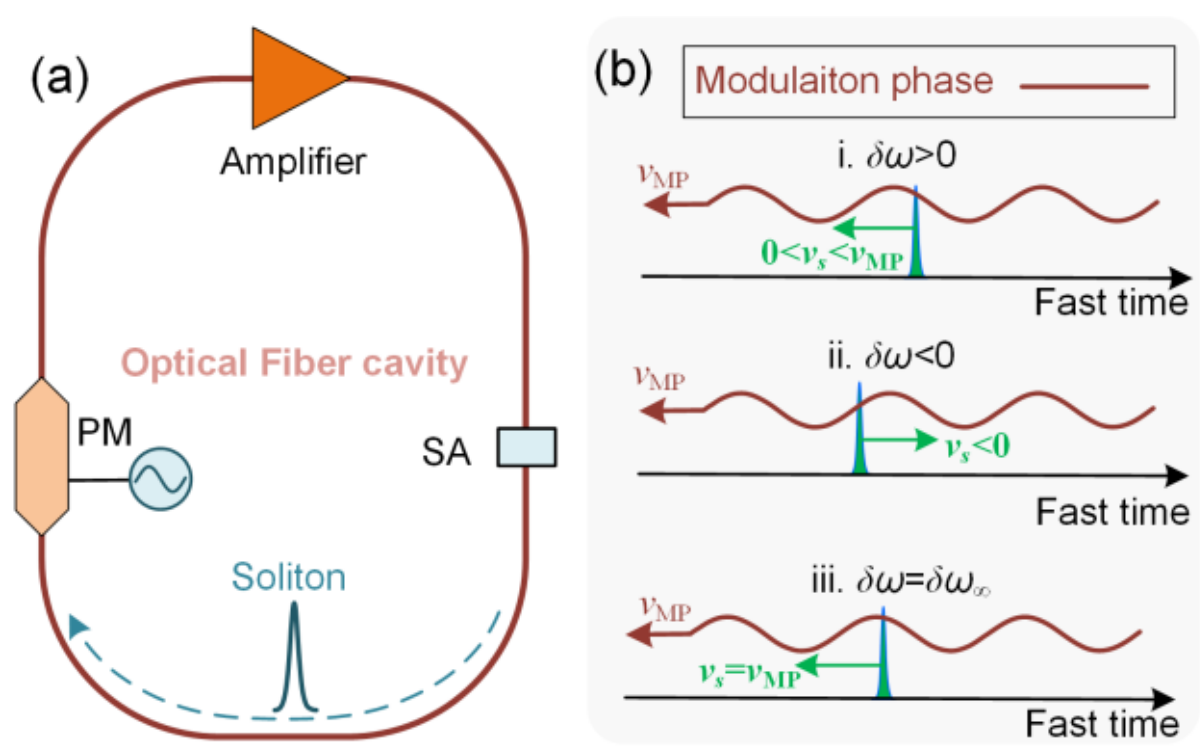

**Figure 1.** (a) Schematic of the mode-locked fiber laser cavity, consisting of an amplifier, a saturable absorber (SA), a phase modulator (PM) driven by an RF signal, and fiber segments. DSs circulate within the cavity. (b) Schematic illustration of a DS (green pulse) interacting with a moving phase potential (red curve). The quantities $v_{MP}$ and $v_s$ correspond to the phase-potential velocity and the DS drift velocity, respectively, both defined relative to the unperturbed DS trajectory. Panels (i) and (ii) correspond to opposite phase gradients, yielding frequency shifts $\delta\omega$ and drift velocities $v_s$ of opposite signs. Panel (iii) corresponds to the synchronization condition $\delta\omega=\delta\omega_\infty$, where $v_s=v_{MP}$ and the DS propagates synchronously with the moving phase potential.

**Figure 1(a)** shows the configuration of the mode-locked fiber laser cavity, which comprises an amplifier, a saturable absorber (SA), a phase modulator (PM), and fiber segments. A phase modulation is imposed on the intracavity field. Figure 1(b) schematically depicts the relative motion between the DS and the moving phase potential. A temporal phase gradient imposed on the pulse induces an instantaneous frequency shift, [43] which is subsequently translated into temporal drift through dispersion. As a result, the DS (green pulse) acquires a drift velocity $v_s$ relative to its unperturbed trajectory in the absence of phase modulation, while the phase potential (red curve) generated by asynchronous phase modulation propagates at a velocity $v_{MP}$.

Three representative dynamical scenarios are illustrated in Figure 1(b-i)–(iii). In Figure 1(b-i) and (ii), the drift cannot fully compensate for the velocity mismatch between the DS and the moving phase potential, causing the DS to repeatedly traverse regions of opposite phase gradient. Consequently, the carrier frequency of the DS undergoes alternating red- and blue-shifts, leading to periodic spectral oscillations. The corresponding frequency shift is quantified by $\delta\omega$, defined as the difference between the instantaneous central frequency of the modulated DS and that of the unmodulated DS (taken as $\delta\omega$=0). The associated variation of the drift velocity further drives the DS to oscillate within the moving phase potential, resulting in temporal meandering. In Figure 1(b-iii), corresponding to the synchronization condition $\delta\omega=\delta\omega_\infty$, the phase-gradient-induced drift exactly compensates the relative motion, satisfying $v_s=v_{MP}$. The DS therefore becomes synchronized with the moving phase potential and remains localized near the potential maximum.

To further investigate this phenomenon, we model the laser system using the extended complex Ginzburg–Landau equation (ECGLE) with an additional phase-modulation term. The normalized ECGLE including the phase-modulation term can be written as:

$$i\frac{\partial\psi}{\partial\xi}-\frac{D}{2}\frac{\partial^2\psi}{\partial\tau^2}+|\psi|^2\psi=R(\xi,\tau), \tag{1a}$$

$$R(\xi,\tau)=i\delta\psi+i\beta\frac{\partial^2\psi}{\partial\tau^2}+i\varepsilon|\psi|^2\psi+i\mu|\psi|^4\psi-J\cos(\omega_M\tau+\Delta\omega\xi)\psi, \tag{1b}$$

where, $\psi=\gamma L_c\psi$ is the normalized envelope of the optical field, $L_c$ is the resonator circumference, $\xi=t/t_R$ is the normalized propagation coordinate, with $t_R=FSR^{-1}$ being the characteristic time scale associated with the resonator mode spacing. $\tau=T/T_0$ is the normalized retarded time in the frame of reference moving with the group velocity and $T_0=\sqrt{|\beta_2|L_c}$ represents the characteristic time scale related to the initial pulse width of the incident light, $D=\mathrm{sgn}(\beta_2)$ is the dispersion propagation regime of propagation, where $\beta_2$ is the group-velocity dispersion coefficient of the resonant cavity. In this work, the overall cavity operates in the anomalous dispersion regime ($\beta_2<0$, i.e., $D=-1$). $\delta$ is the average linear gain coefficient, $\beta$ describes spectral filtering (gain dispersion) effect, and $\varepsilon$ and $\mu$ correspond to nonlinear gain and its higher-order saturation effect, respectively. The final term in **Equation. (1b)** accounts for the effect of the high-frequency EOM. The phase modulation parameters $J$, $\omega_M=2\pi T_0 f_M$ and $\Delta\omega=2\pi t_R\Delta f$ denote the modulation amplitude, modulation angular frequency, and detuning angular frequency, respectively.

The ECGLE describes the pulse dynamics in phase modulated mode-locked lasers, where the interplay between dispersion and Kerr nonlinearity sustains the underlying soliton-like pulse profile. The term $R(\xi,\tau)$ on the right-hand side incorporates dissipative effects, higher-order nonlinearities, and phase-modulation term, which are treated as perturbative contributions in the present analysis. To analytically characterize the DS dynamics, we construct a reduced model via a perturbative variational approach. This method was first formulated for optical fiber solitons by Anderson in his seminal work, and was later generalized and developed for dispersion-managed soliton systems in subsequent studies. [44,45] We adopt the following DS envelope ansatz $\psi(\xi,\tau)=B\mathrm{sech}\left[(\tau-\langle\tau\rangle)/\tau_s\right]\mathrm{e}^{-i\langle\omega\rangle(\tau-\langle\tau\rangle)}\mathrm{e}^{i\varphi}$, where $\tau_s$ denotes the pulse duration and $B$ represents the pulse amplitude. Since the overall cavity operates in the normal dispersion regime, the balance between dispersion, nonlinearity, and dissipative effects enforces a fixed relation between DS amplitude and width, satisfying $B\tau_s=1$ under the present normalization scheme. [46] By applying the variational procedure to the perturbed system, [39,41,44,45] we derive the evolution equations governing the DS central frequency and temporal position:

$$\frac{\mathrm{d}\langle\omega_0\rangle}{\mathrm{d}\xi}=-\frac{4\beta\langle\omega_0\rangle}{3\tau_s^2}+J\omega_M\sin(\omega_M\langle\tau\rangle+\Delta\omega\xi),$$
$$\frac{\mathrm{d}\langle\tau_0\rangle}{\mathrm{d}\xi}=D\langle\omega_0\rangle. \tag{2}$$

The evolution equations for the phase $\varphi$ and amplitude $B$ are not relevant to the following analysis and are therefore omitted. From the variational results, we find that in the locked state, where the oscillation period diverges, the detuning angular frequency is $\Delta\omega_\infty=3J\tau_s^2\omega_M^2/4\beta$, at which the spectral amplitude attains its maximum value $A_\infty=3J\tau_s^2\omega_M/4\beta$. Meanwhile, fitting formulas for the DS oscillation period and spectral oscillation amplitude are given:

$$T_{\mathrm{O}} = 2\pi \cdot \left(\Delta\omega^2 - \Delta\omega_\infty^2\right)^{-1/2},$$
$$A = A_l \sin\left(\left[\left(2/\pi\right)^2 + \left(A_l/A_\infty\right)^2\right]^{-1/2}\right). \tag{3}$$

### 3. Detuning-Dependent Spectral–Temporal Dynamics

Numerical simulations based on **Equation (1)** are performed to investigate the dynamics of DSs in a moving phase potential, and the results are presented in **Figure 2**. The detuning induces a relative motion between the DS and the phase potential. At sufficiently large detuning, the DS cannot continuously track the moving potential and therefore repeatedly traverses regions of opposite phase gradient. As a result, alternating red- and blue-shifts of the DS carrier frequency occur, leading to periodic spectral oscillations.

Figure 2(a) and (b) show the spectral and temporal evolutions of the DS at a detuning angular frequency of 0.024, respectively. The color maps show the numerical simulation results. Pronounced periodic spectral oscillations are observed are observed in response to the moving phase potential. The green dashed line in Figure 2(a) indicates the spectral centroid. Using the centroid evolution equation obtained from the variational analysis **Equation (2)**, the temporal trajectory can be reconstructed through $\langle\tau\rangle(\xi) = D\int_0^{\xi} \langle\omega\rangle_{\xi}\, d\xi$. The reconstructed temporal trajectory, shown as the green curve in Figure 2(b), reproduces the temporal evolution of the DS remarkably well. This agreement confirms that the temporal meandering of the DS arises from the oscillatory frequency shifts.

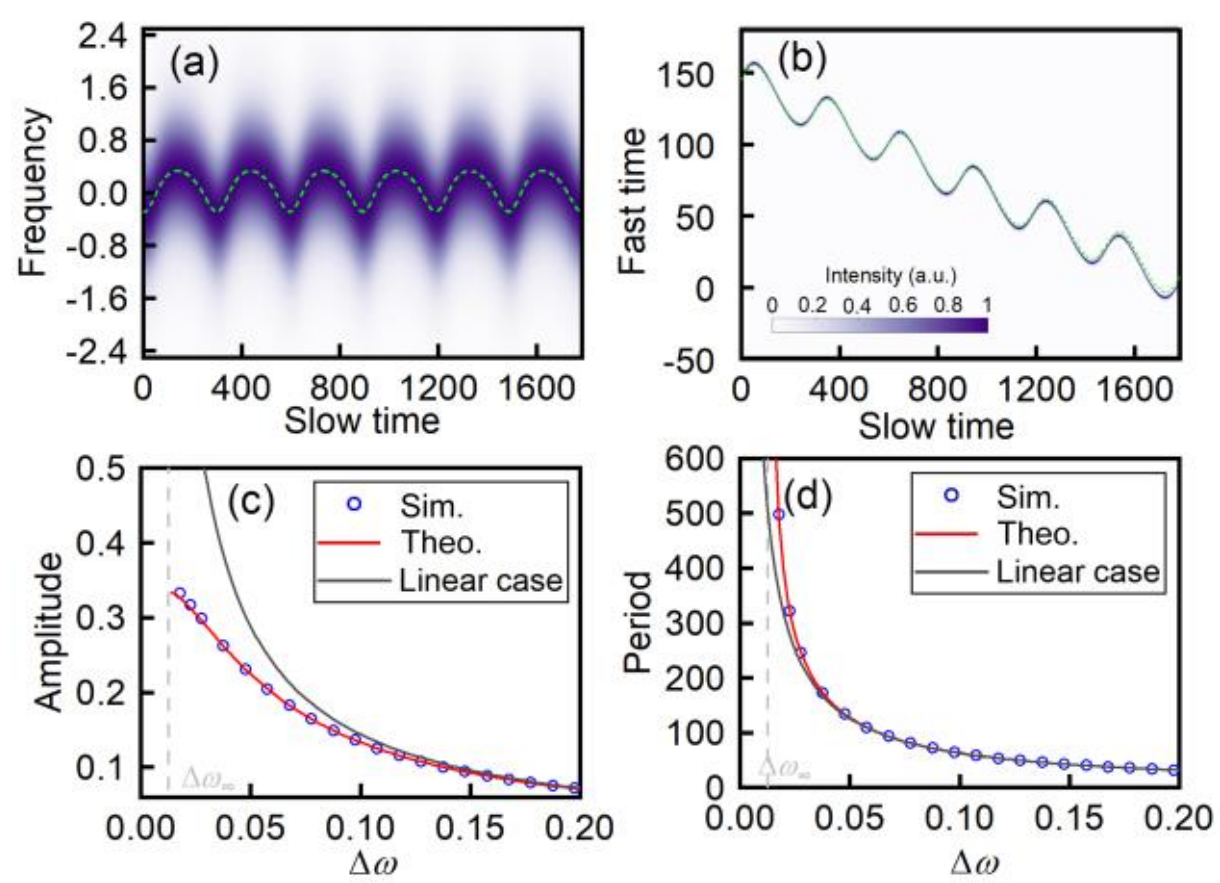

**Figure 2.** Numerical simulation results. (a) Spectral evolution of the DS at a detuning angular frequency $\Delta\omega = 0.024$. (b) Corresponding temporal evolution. The colormaps show the numerical simulation results, and the green dashed lines indicate the evolution of the spectral centroid and temporal positions of DS, respectively. (c) Spectral oscillation amplitude $A$ and (d) oscillation period $T_{\mathrm{O}}$ as functions of detuning angular frequency. Blue circles denote numerical results, red solid lines are theoretical predictions of Equation (2), and black solid lines indicate the ideal linear predictions. The gray dashed line marks the synchronization threshold $\Delta\omega_\infty$=0.0193.

Furthermore, Figure 2(c) and (d) present the dependence of the spectral oscillation amplitude and oscillation period as functions of the detuning angular frequency. The blue circles represent the numerical results. The black solid lines correspond to the linear predictions obtained by neglecting the dissipative term, and the red curves denote the theoretical predictions derived from Equation (2). The corresponding threshold $\Delta\omega_\infty$ is marked by the gray dashed line. Excellent agreement between the numerical simulations and the variational theory confirms the validity of the theoretical description over the entire detuning range.

As the detuning decreases and approaches the threshold angular frequency $\Delta\omega_\infty$, the oscillation amplitude and period exhibit markedly different behaviors. The linear model predicts that the oscillation amplitude diverges at the threshold. In contrast, both the numerical simulations and the variational theory show that the amplitude remains finite as the locking threshold is approached and reaches a maximum value of $A_\infty$=0.54 at $\Delta\omega_\infty$=0.0193. This discrepancy indicates that dissipative effects, neglected in the linear approximation, become increasingly important near the locking threshold and prevent the unbounded growth predicted by the lossless model. Unlike the oscillation amplitude, the oscillation period increases continuously as the detuning approaches $\Delta\omega_\infty$ and diverges at the threshold, in agreement with both the variational theory and numerical simulations. This divergence indicates that the oscillatory dynamics become progressively slower as the threshold is approached. At the threshold, the oscillatory state disappears and the DS becomes locked to the moving potential.

For larger detuning, both the oscillation amplitude and period gradually approach the corresponding linear predictions. This trend indicates that the influence of dissipative effects becomes progressively weaker, and the dynamics approach the linear limit. The consistency between the numerical simulations and the theoretical predictions confirms the validity of the variational model. In particular, the model accurately reproduces the dependence of both the oscillation amplitude and period on the detuning and correctly predicts the synchronization threshold.

## 4.Experimental Setup

As shown in **Figure 3**, the proposed scheme is experimentally implemented in a mode-locked fiber laser an intracavity PM. A 6-m-long erbium-doped fiber (EDF) is pumped by a 976-nm laser diode through a wavelength-division multiplexer (WDM). A polarization-insensitive isolator (PI-ISO) ensures unidirectional propagation of the intracavity field, while a 90:10 optical coupler (OC) extracts the pulse train from the laser cavity. A carbon-nanotube saturable absorber (CNT-SA) is employed to initiate and sustain mode-locking operation. The PM is driven by a radio-frequency (RF) signal generator and provides a phase-modulation coefficient of 0.35 rad/$V_{pp}$, thereby imposing a periodic phase modulation on the intracavity field. All fiber pigtails are standard single-mode fibers (SMFs).

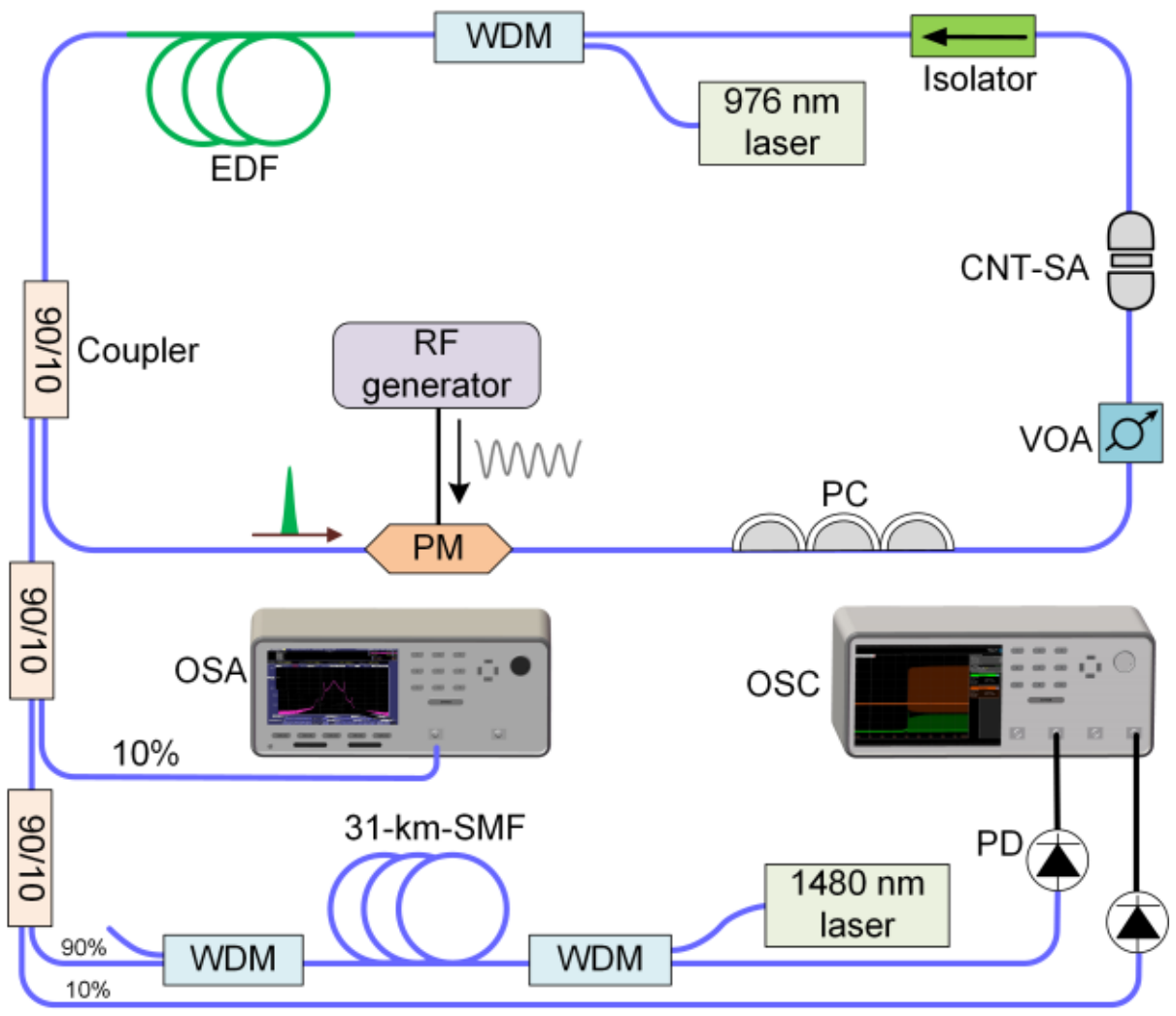

**Figure 3.** Detailed schematic of the experimental setup. EDF, erbium-doped fiber; WDM, wavelength-division multiplexer; CNT-SA, carbon-nanotube saturable absorber; VOA, Adjustable optical attenuator; PC, polarization controller; OSA, optical spectrum analyzer; OSC, oscilloscope; PD, photodetector.

The total cavity length is 17.6 m, corresponding to a fundamental repetition rate of 11.74 MHz. The net cavity dispersion is -0.022 ps$^2$/m. The laser output is extracted through the 90/10 OC and divided into three measurement channels. One channel is connected to an optical spectrum analyzer (OSA). A second channel is directly detected by a photodetector (50 GHz bandwidth) and monitored with a real-time oscilloscope with a bandwidth of 33 GHz to monitor the temporal evolution of the DS. The third channel is used to record the spectral evolutions of optical pulses via the dispersive Fourier transform (DFT) technique. [47-51] In this measurement, the pulses are temporally stretched by propagation through a 31-km spool of SMF, and subsequently recorded by a high-bandwidth real-time oscilloscope (OSC, 33 GHz bandwidth).

## 5. Experimental Validation and Dynamical Tunability

The experimental results are presented in **Figure 4**, where the DS exhibits spectral-temporal dynamics under phase modulation. The spectral evolution maps shown in Figure 4(a), 4(c), and 4(e) reveal periodic spectral oscillations of the DS. This behavior is closely related to the relative position of the DS within the moving phase potential, consistent with the physical mechanism established in our preceding theoretical and numerical analyses. The green dashed curves denote the spectral centroid of the DS. The corresponding temporal evolutions are shown in Figure 4(b), 4(d), and 4(f). In addition to the oscillations, the DS exhibits a pronounced temporal drift during its evolution. The green curves in Figure 4(b), 4(d), and 4(f) are obtained by integrating the spectral centroid evolution in Figure 4(a), 4(c), and 4(e), respectively, using the same method as described in Figure 2(b). The calculated trajectories agree well with the experimentally measured temporal evolution, indicating that the temporal drift originates from the frequency shifts through dispersion. From a macroscopic perspective, the temporal meandering can be understood as a consequence of the continuous pulling exerted by the moving phase potential.

Figure 4(g) and 4(h) summarize the experimentally measured spectral oscillation amplitude and period as functions of the detuning frequency, respectively. The blue circles denote the experimental data. The red curves are obtained from the fitting expressions in **Equation (3)**, and the black solid lines represent the corresponding linear predictions. The corresponding locking threshold $\Delta f_\infty$ is indicated by the gray dashed line.

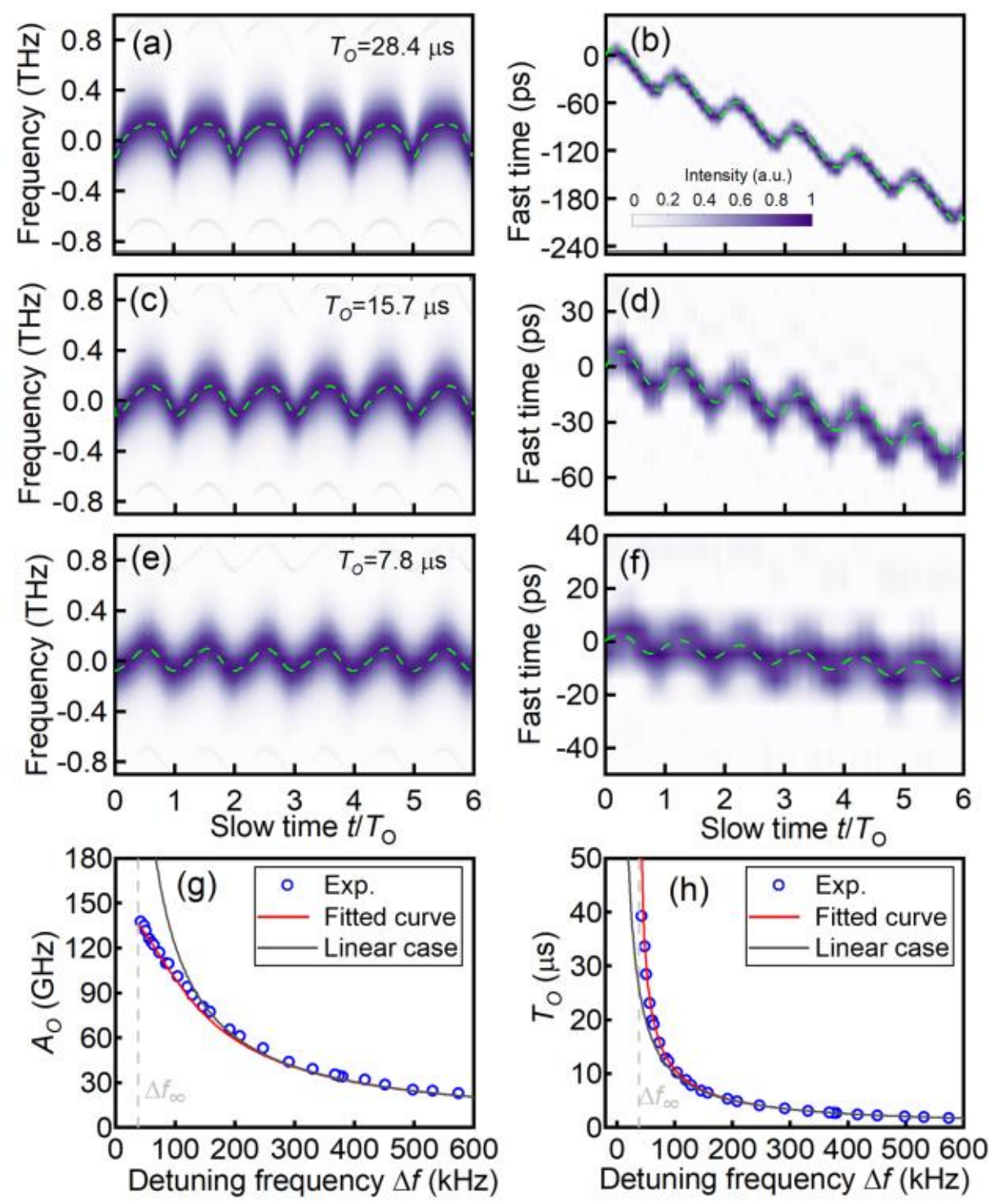

**Figure 4.** Experimental results. (a), (c), and (e) Spectral evolution of DS at oscillation periods $T_O$=28.4 μs, $T_O$=15.7 μs and $T_O$=7.8 μs, respectively. The colormaps are experimental results, and the green dashed lines indicate the evolution of the DS spectral centroid. (b), (d), and (f) Corresponding temporal evolution. (g) Measured oscillation amplitude $A$ and (h) oscillation period $T_O$ as functions of detuning frequency. Blue circles denote the experimental results, red solid lines represent fitted results from Equation (3), and black solid lines indicate the corresponding linear predictions. The gray dashed line marks the experimentally observed threshold $\Delta f_\infty$=36 kHz.

A clear detuning dependence is observed in both the spectral oscillation amplitude and the period. In the low-detuning regime, the measured amplitude deviates markedly from the linear prediction, indicating that dissipative effects remain important and suppress its growth. Experimentally, the oscillation amplitude reaches a maximum value of approximately 139 GHz at $\Delta f_\infty$=36 kHz, corresponding to the locking threshold. Meanwhile, the oscillation period increases rapidly as the detuning decreases, revealing a strong slowing of the dynamics. As the detuning increases, both the oscillation amplitude and period gradually approach the linear predictions, indicating that the influence of dissipative effects becomes less pronounced and the dynamics progressively approach the linear-response regime. The fitted curves closely follow the experimental data over the full detuning range, confirming that Equation (3) provides an accurate description of the observed dynamics. These measurements therefore demonstrate that the detuning not only controls the amplitude and timescale of the DS oscillations, but also governs the transition from dissipation-dominated dynamics toward the linear-response regime.

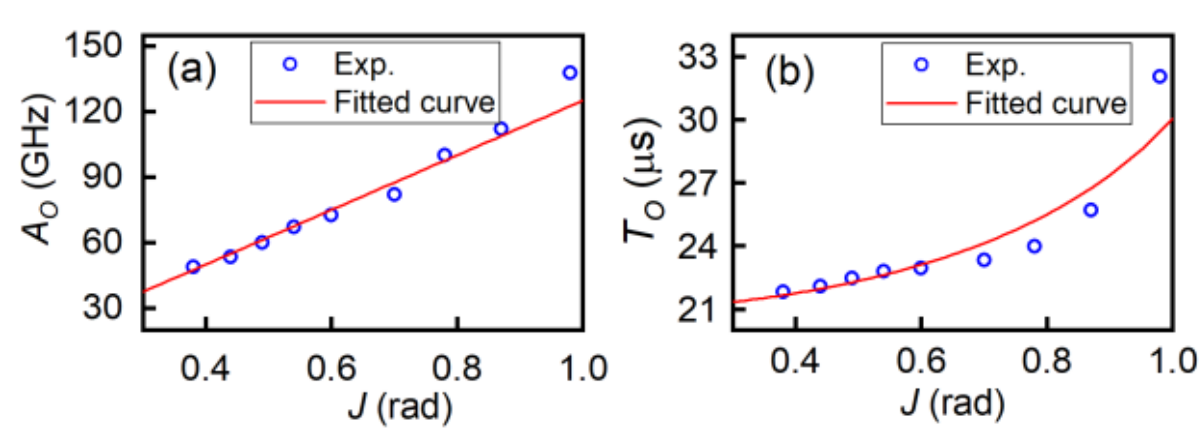

**Figure 5.** Experimental results. (a) Spectral oscillation amplitude $A$ and (b) oscillation period $T_O$ as functions of the modulation depth $J$. Blue circles denote the experimental results, and the red solid lines represent fitted results from Equation (3).

**Figure 5** shows the dependence of the DS oscillation dynamics on the modulation depth. The blue circles in Figure 5(a) and 5(b) summarize the dependence of the oscillation amplitude and oscillation period on the modulation depth, respectively. As shown in Figure 5(a) and 5(b), the fitted curves based on Equation (3) closely follow the experimental results. The results in Figure 5(a) indicate that the modulation depth enables control of both the spectral oscillation amplitude and the maximum frequency shift $A_\infty = 3J\tau_s^2\omega_M/4\beta$ under locking. Furthermore, the locking threshold $\Delta\omega_\infty = 3J\tau_s^2\omega_M^2/4\beta$ increases with modulation depth, implying an expansion of the synchronization range. Figure 5(b) shows that the oscillation period becomes longer as the modulation depth increases.

These observations can be attributed to the interaction between the DS and the phase-induced potential. Increasing the modulation depth enhances the amplitude of the intracavity phase potential and thereby strengthens the phase-gradient-induced driving force acting on the DS. As a result, both the spectral oscillation amplitude and the maximum frequency excursion at the locking threshold increase with modulation depth. Meanwhile, the stronger phase potential modifies the oscillatory dynamics of the DS, leading to an increase in the oscillation period.

Throughout the experiments, the DS remained in a stable single-pulse state without exhibiting pulse splitting, or other dynamical instabilities. The smooth dependence of the measured parameters on the modulation depth further indicates a broad parameter range for stable oscillations in the phase-modulated cavity. The detuning and modulation depth play distinct roles in the dynamics: the former governs the synchronization condition between the modulation and the cavity repetition rate, whereas the latter determines the strength of the external phase perturbation. The combined tuning of these two parameters enables systematic control of multiple dynamical properties, including the oscillation amplitude, oscillation period, and synchronization of the DS.

## 6.Discussion and Conclusion

The present study combines numerical simulations, experimental observations, and theoretical analysis to investigate the spectral–temporal dynamics of dissipative solitons in a moving intracavity phase potential generated by high-frequency asynchronous phase modulation. The results demonstrate that the drifting phase potential induces robust periodic spectral oscillations and temporal meandering. At large detuning, the dissipative soliton fails to continuously track the moving potential and experiences alternating phase gradients, resulting in periodic red- and blue-shifts of the central frequency.

As the detuning approaches a critical value, the oscillation period increases and diverges, indicating a continuous transition from a nonstationary oscillatory regime to a locked state with a finite frequency shift. The modulation depth further controls the oscillation amplitude and period, providing an additional degree of freedom for tuning the dynamics. More generally, these results establish a clear dynamical picture of dissipative soliton evolution in moving external potentials and highlight the role of asynchronous driving in shaping nonequilibrium states in nonlinear optical cavities.

Throughout the parameter space explored, the soliton remains stable as a single-pulse state without pulse splitting or multi-soliton competition, indicating a robust dynamical regime. From an application perspective, the frequency-shifted locked state provides a potential route toward electrically controlled spectral tuning in mode-locked fiber lasers. By jointly tuning the modulation depth and detuning, substantial and deterministic frequency shifts can be achieved without mechanical tuning elements, consistent with the theoretically predicted locked offset $A_\infty$.

**Acknowledgements**

This work was supported by National Natural Science Foundation of China (NSFC) (Grant No. 12475006, 12247103, and 12434006).

**Conflicts of Interest**

The authors declare no conflicts of interest.

**Data Availability Statement**

The data that support the findings of this study are available from the corresponding author upon reasonable request.

## Spectral Oscillations and Temporal Meandering of Dissipative Solitons in a Moving Intracavity Phase Potential

Asynchronous phase modulation creates a moving intracavity phase potential that drives dissipative solitons (DSs) out of equilibrium. Under detuning, the DS cannot follow the moving potential, leading to a timescale mismatch. This mismatch gives rise to spectral oscillations and temporal meandering, while sufficiently small detuning yields synchronized locking. The oscillatory dynamics are controlled by modulation detuning and modulation depth.

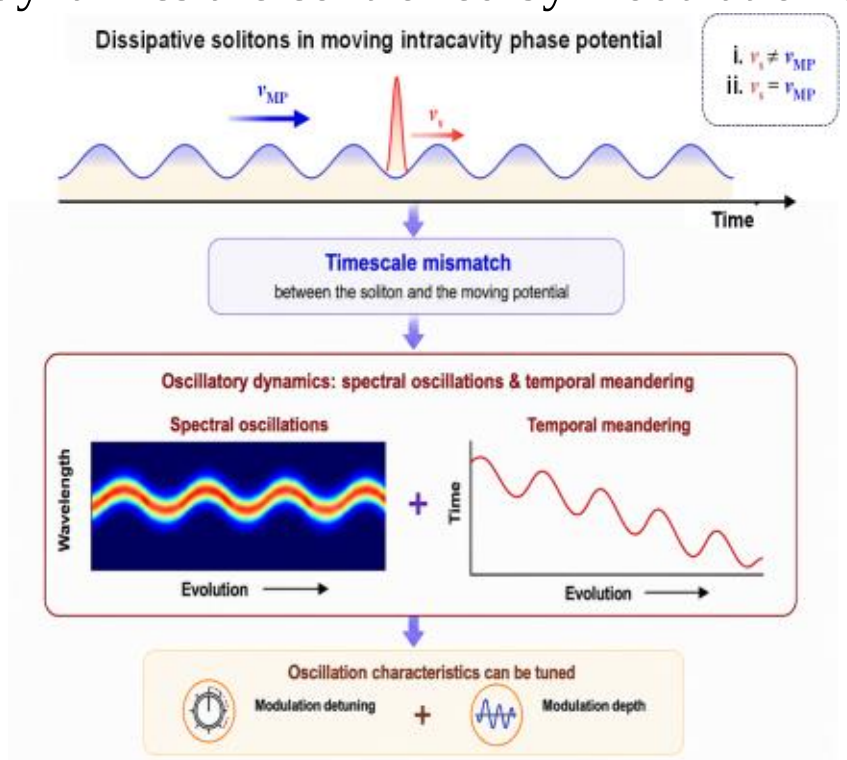